\documentclass[RNAAS]{aastex62}

\begin{document}

\title{Non-Detection of a Helium Exosphere for the Hot Jupiter WASP-12\lowercase{b}}

\author{Laura Kreidberg}
\affiliation{Harvard-Smithsonian Center for Astrophysics, 60 Garden Street, Cambridge, MA 02138}
\affiliation{Harvard Society of Fellows, 78 Mount Auburn Street, Cambridge, MA 02138}

\author{Antonija Oklop\v{c}i\'{c}}
\affiliation{Harvard-Smithsonian Center for Astrophysics, 60 Garden Street, Cambridge, MA 02138}

\section{Introduction}
An exosphere was recently detected around the exoplanet WASP-107b, a
low-density, warm Neptune \citep{spake18}, based on absorption features from metastable helium at $10833\,\mathrm{\AA}$ \citep[predicted
by][]{seager00,oklopcic18}. The helium feature provides a new probe of atmospheric escape that is
advantageous in several ways: (1) it is observable with near-infrared facilities
(in contrast to most other signposts of atmospheric escape that appear in the
ultraviolet) and (2) it is minimally affected by interstellar absorption,
thereby opening the door to studying atmospheric escape in a greater number of systems.

Inspired by the WASP-107b detection, we searched archival HST observations of
another evaporating exoplanet, WASP-12b, for signs of a helium exosphere.
WASP-12b is a promising 
candidate for this search because it is one of the hottest known hot
Jupiters \citep[$T_\mathrm{eq} = 2500$ K;][]{hebb09}. At this level of intense
irradiation, theory predicts a high rate of escaping atoms and molecules from
the planet's atmosphere, and indeed, transit observations in the ultraviolet
have revealed a patchy cloud of escaping material \citep{nichols15, salz16}.  

\section{Observations and Data Reduction}
\begin{figure*}
\begin{centering}
\includegraphics[width = 0.9\textwidth]{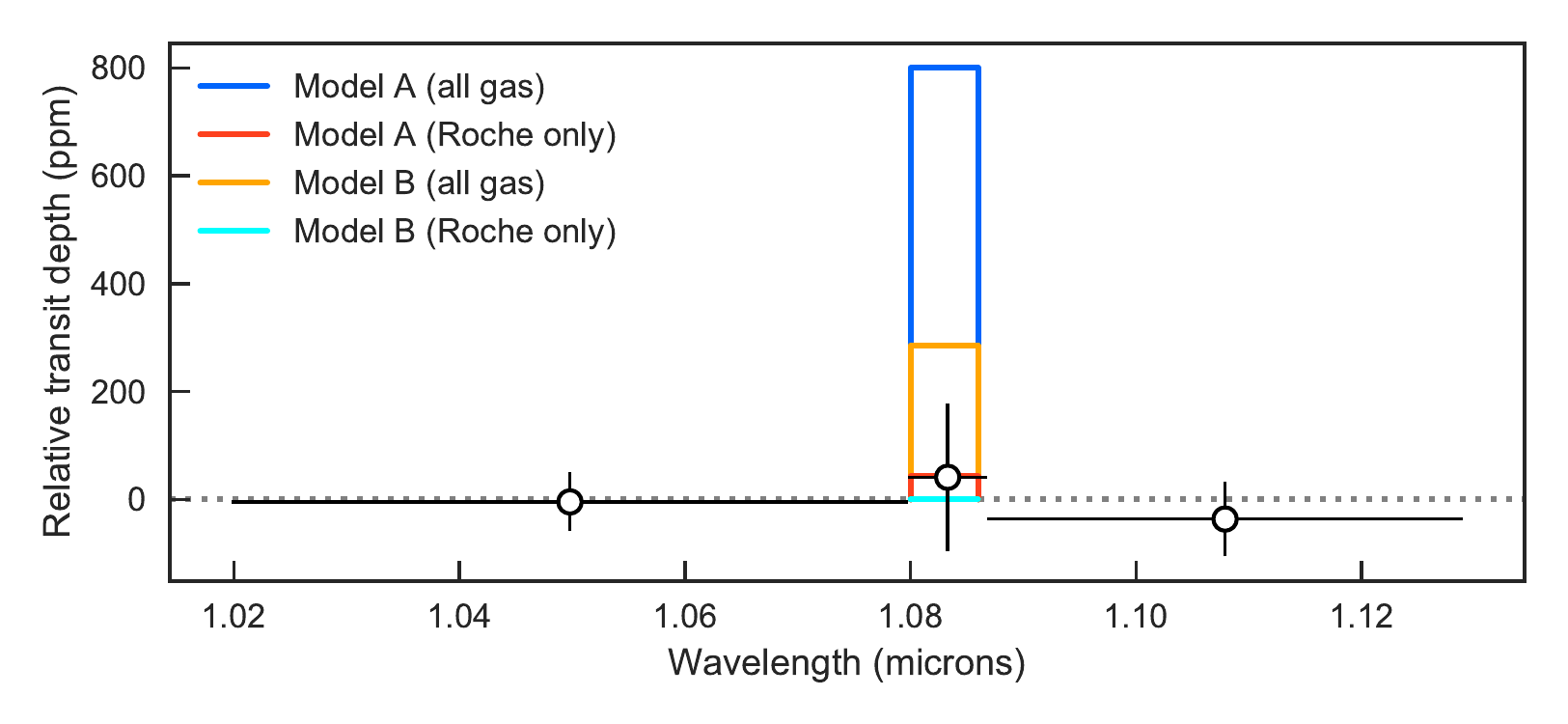}
\caption{Transmission spectrum of WASP-12b, compared to model predictions for
the strength of the $10833\mathrm{\AA}$ helium feature.}
\end{centering}
\label{fig:spectrum}
\end{figure*}

For this Note, we reanalyzed three transits of WASP-12b observed with the
Hubble Space Telescope/Wide Field Camera 3 G102 grism.  In our analysis, we used the same methodology as
\cite{kreidberg15b}, except with different spectral binning to include a narrow
band ($70\,\mathrm{\AA}$, the spectrograph's native resolution) centered on the
helium feature, with two wider bands at adjacent wavelengths. 
The transmission spectrum (shown in Figure\,\ref{fig:spectrum}) is consistent with that reported in
\cite{kreidberg15b} and shows no evidence for variability between epochs.  Surprisingly, there
is no significant increase in transit depth at $10833\,\mathrm{\AA}$. The
transit depth for the helium feature is just $59 \pm 143$ ppm larger than the
weighted mean depth in the adjacent wavelength bins, in contrast to past
observations of the exosphere in the NUV, which show an increase of $\sim1\%$
relative to the optical transit depth \citep{nichols15}.

\section{Model Predictions}
To predict the helium absorption signal of WASP-12b,
we used the theoretical model described in \cite{oklopcic18}.
In this 1D model, we assumed the thermosphere of the planet is composed of atomic hydrogen and helium in 9:1 number
ratio. For the thermospheric density and velocity profiles we adopted the
isothermal Parker wind model, assuming a gas temperature of $T=10^4$~K and a  
total atmospheric mass loss rate of $4\times 10^{11}$~g~s$^{-1}$ \citep[based on
hydrodynamic simulations of atmospheric escape in WASP-12b
from][]{salz16}. We used the solar irradiance spectrum as the input spectrum. 
We considered two density profiles of metastable helium: one calculated for the
gas at the substellar point, and the other for the
gas at the terminator, which is likely a better approximation of the 3D
gas distribution. We label these scenarios Model A and B,
respectively. For each profile, we predict the expected absorption signal for: 1) gas within the Roche radius of the planet, where the assumptions of our 1D model are more likely valid, and 2) all the gas out to 20 planetary radii. 
As illustrated in Figure\,\ref{fig:spectrum}, the predicted helium feature
amplitudes are generally small and agree well with the observations (with the
exception of the least physically plausible scenario, Model A with all gas,
which is inconsistent with the observed spectrum at $5.2\,\sigma$ confidence). 


\section{Discussion}
This non-detection raises the question of why the helium feature is larger in  
WASP-107b than WASP-12b, which also has an exosphere.  Even though
WASP-12b experiences 20$\times$ higher bolometric flux, there are several factors that may contribute to a
larger signal for WASP-107b. First, the Roche radius of WASP-107b relative to
its host star is $2\times$ larger than for WASP-12b. In addition, the incident stellar spectrum for WASP-107b
is more favorable for producing metastable helium.  WASP-107 is an active star
and expected to be bright in the extreme ultraviolet (EUV), which is responsible for populating the excited metastable
state. By contrast, common stellar activity indicators suggest that WASP-12
is an unusually inactive star, and hence it might be even fainter in the EUV
than a typical Sun-like star.  In addition, the later spectral type of WASP-107 (K6 versus G0) results
in a lower flux of hydrogen-ionizing radiation, which reduces the density of
free electrons (collisions with electrons are the main de-populating mechanism
of the metastable state).  Finally, the escape rate of material from WASP-12b is
so high that it may produce a torus of material around the star
\citep{haswell17, debrecht18}. If the metastable helium is distributed uniformly
around the star, the helium absorption feature would exist at all orbital phases, not just during the planet's transit.

In conclusion, metastable helium remains a promising probe of atmospheric
escape, but the amplitude of the signal is highly sensitive to the input stellar
spectrum and the geometry of
the evaporating gas cloud.  These considerations should be taken into account in the design of future searches for helium exospheres.

\acknowledgements
We thank Caroline Morley and Hannah Diamond-Lowe for helpful discussions at House of Chang.


 \newcommand{\noop}[1]{}

\end{document}